\documentclass[conference]{IEEEtran}
\IEEEoverridecommandlockouts
\usepackage{cite}
\usepackage{enumitem}
\usepackage{amsmath,amssymb,amsfonts, amsthm}
\usepackage[font=small]{caption}
\usepackage{graphicx}
\usepackage{textcomp}
\usepackage{xcolor}
\def\BibTeX{{\rm B\kern-.05em{\sc i\kern-.025em b}\kern-.08em
		T\kern-.1667em\lower.7ex\hbox{E}\kern-.125emX}}

\usepackage{algorithm}
\usepackage[noend]{algpseudocode}

\begin{document}
	
	\title{Performance Effectiveness of Multimedia Information Search Using the Epsilon-Greedy Algorithm}
	
	\author{\IEEEauthorblockN{Nikki Lijing Kuang}
		\IEEEauthorblockA{\textit{Department of Computer Science and Engineering} \\
			\textit{University of California San Diego}\\
			La Jolla, USA \\
			l1kuang@ucsd.edu}
		\and
		\IEEEauthorblockN{Clement H.C. Leung}
		\IEEEauthorblockA{\textit{School of Science and Engineering} \\
			\textit{The Chinese University of Hong Kong}\\
			Shenzhen, China \\
			clementleung@cuhk.edu.cn}
	}
	
	\maketitle
	
	\begin{abstract}
		In the search and retrieval of multimedia objects, it is impractical to either manually or automatically extract the contents for indexing since most of the multimedia contents are not machine extractable, while manual extraction tends to be highly laborious and time-consuming. However, by systematically capturing and analyzing the feedback patterns of human users, vital information concerning the multimedia contents can be harvested for effective indexing and subsequent search. By learning from the human judgment and mental evaluation of users, effective search indices can be gradually developed and built up, and subsequently be exploited to find the most relevant multimedia objects. To avoid hovering around a local maximum, we apply the $\epsilon$-greedy method to systematically explore the search space. Through such methodic exploration, we show that the proposed approach is able to guarantee that the most relevant objects can always be discovered, even though initially it may have been overlooked or not regarded as relevant. The search behavior of the present approach is quantitatively analyzed, and closed-form expressions are obtained for the performance of two variants of the $\epsilon$-greedy algorithm, namely EGSE-A and EGSE-B. Simulations and experiments on real data set have been performed which show good agreement with the theoretical findings. The present method is able to leverage exploration in an effective way to significantly raise the performance of multimedia information search, and enables the certain discovery of relevant objects which may be otherwise undiscoverable.
	\end{abstract}
	
	
	\section{Introduction}
	Effective search and retrieval of multimedia content has long been a challenging task, owing to the intrinsic complexity caused by the diversified heterogeneous modalities. The difficulty is further magnified by the increasingly expressive and intelligent searches that users attempt to perform. Compared to the conventional unambiguous query terms, such queries tend to be heuristic and multi-semantic in the pattern of natural languages, where contextual and implicit information can often play a significant part. Examples can be, "a song of jazz pop without saxophone" and "images of superbloom in California with humans wearing hats". The presence of \textit{user intent} \cite{kofler2016user, xie2019user, liem2017multimedia} therefore widens even more the semantic gap between low-level features and high-level semantics \cite{cao2017transitive, azzam2004implicit, stevenson2005comparative} in multimedia information retrieval (MIR). 
	
	To address the semantic gap in MIR, indexing technique empowered by user feedback patterns has led to a different yet promising direction \cite{datta2017multimodal, leung2012intelligent}. Unlike content-based information retrieval, indexing in MIR is typically used with available textual metadata or implicit user information for efficient access. However, existing algorithms typically aim to retrieve multimedia objects from the search space purely based on their relevance to the proposed queries. With measures such as \textit{tf-idf}, objects indicated to be the most relevant are returned to form search results. The initial indexing thus plays a crucial role in the evolution of presented search results. As objects are ranked according to the index scores, those with initial large values tend to dominate the presented list, making it nearly impossible to discover hidden objects that are potentially of greater relevance. Meanwhile, the problem of local optimum also results in the failure to adapt search results to the latest user interests along with time.
	
	In this paper, we move one step forward to address the aforementioned problems by balancing exploitation and exploration. The dilemma of exploitation and exploration is ubiquitous and has been widely studied in reinforcement learning (RL) and control systems \cite{fruit2018near, narayanan2017event, vignesh2017online}. Here, we are interested in the effective exploration in the context of index-based multimedia search that encapsulates mental judgment from users to capture high-level linguistic semantics. By learning from implicit user feedback, search indexes are built up to reflect the latest user intent and information needs, which can then be further exploited to uncover the most relevant objects. We aim to enable possible discovery of hidden relevant objects, where unfavorable initial indexes are bound with them, resulting in the failure of roll-out for selection. Specifically, we are interested in the worst case scenario that for a specific query, the ground-truth of most relevant objects are initially hidden due to inadequacies in initial indexes. By incorporating the $\epsilon$-greedy algorithm \cite{sutton2018reinforcement} with index-based search methods, proactive exploration in the search space becomes possible, ensuring the exposure of objects of interest. In the meantime, current information is exploited to present satisfactory search results. Unlike entirely random methods that do not harness any useful information, we stress on the systematic evolution that achieves a balance between exploitation and exploration.
	
	The main contributions of this work are as follows.
	\begin{itemize}
		\item We explore the problem of how to effectively perform exploration in index-based multimedia search methods. The goal here is not to devise new methodologies for MIR, but to instead balance exploitation and exploration within a widely-adopted index based frameworks, so as to provide a \textit{theoretically justified} approach for effective exploration. Specifically, implicit user feedback patterns are learned and harnessed to form semantic indexes, while the $\epsilon$-greedy algorithm is incorporated for discovering hidden relevant multimedia objects.
		\item Two variants of the $\epsilon$-greedy algorithms, EGSE-A and EGSE-B, are presented for exploration of hidden relevant objects. Analysis is performed to examine the worst case scenario where the evolution pattern of the most relevant objects start with unfavorable initial indexes. Closed-form expressions are obtained with detailed performance evaluation, which guarantee the discovery of interested objects in finite time. 
		\item To validate and corroborate the effectiveness of exploration and exploitation, Monte-Carlo simulations and experiments on real data set are performed. 
	\end{itemize}
	
	The organization of this paper is as follows. Section 2 compares the present study and recent works. Preliminaries and problem formulation are given in Section 3. Section 4 analyzes the performance of the algorithms, followed by experimental evaluations in Section 5. Finally, Section 6 concludes the work.
	
	\section{Related Works}
	Relevance feedback in various forms \cite{zamani2017relevance, peng2017overview, Joachims2017, dang2017multimodal, over2004multimedia} have been used to effect improvements in search performance \cite{sarwar2018term, singhi2014cultural, kofler2015uploader, mei2014multimedia, cui2014social}. Applying an evolutionary method for classifying images and building it into the search architecture is studied in \cite{real2018regularized}. Related search problems employing some form of evolutionary behavior are also proposed in \cite{liu2017hierarchical, liu2018darts, rashedi2018comprehensive, franzoni2017semantic, hirsch2017document}. Using click information for relevance feedback have been employed in \cite{lu2014inferring}, and \cite{pan2014click}. In \cite{lu2014inferring}, it infers user image search goals from the users’ response. While it makes use of click information similar to our present study, it also relies on image analysis by extracting the color, texture, and shape features to produce a feature vector for each image. Unlike our method, the systematic exploration of the search space is impractical due to its substantial computational overhead. An interesting cross-view method of learning based primarily on click-through is employed for image search in \cite{pan2014click}. Different from the present method, the training mechanism of cross-view learning is carried out by minimizing an objective function representing the distance between query and image mappings, while maintaining the relationships between the training examples in the original feature space. In comparison with our method, the approach there will incur substantial training and optimization overheads, which can become computationally expensive.
	
	Combining search with learning mechanisms are increasingly recognized as useful and have been proposed in \cite{gao2017learning, zheng2018discriminatively, yang2015click, li2018learning, zhang2008topological}. In \cite{gao2017learning}, the authors introduce a framework using feature selection, encoding, and learning, and apply these to retrieval and annotation. While the present study addresses the similar problems of multimedia retrieval and annotation (since indexing is a form of annotation), instead of using a purely machine-based learning approach, we exploit human expertise and use them as agents for exploration, so that the entire search space may be covered. The net result is that a precision rate close to 100\% may be attained. Similarly, learning is applied to person identification and generalized to image retrieval in \cite{zheng2018discriminatively}. While that study makes use of CNN to obtain descriptors of pedestrians, our study learns directly from human inputs without requiring extensive training. In \cite{yang2015click}, the problem of ranking is considered, and a re-ranking mechanism, called click-boosting multi-modality graph-based re-ranking, is proposed. That algorithm makes use of clicked images to locate similar images, and re-ranks them. Given a query, an optimization problem needs to be solved. The present study is different from this as algorithmically solving an optimization is not required, nor is ranking considered to be important in the search results. Whilst in \cite{li2018learning}, a label preserving multimedia hashing learning framework is presented, which learns the associated codes by solving a series of integer programming problems. Unlike that study which makes use of hash functions for search identification, we tackle the problem directly from the intent of the human users and learn from their interaction with the presented results, which has been increasingly recognized as an important element in multimedia search \cite{kofler2016user, ellis2014predicting, riegler2014reflects}.
	
	The value of information captured through human behavior has been recognized in \cite{leung2012intelligent, cobarzan2017interactive} and \cite{park2015large}. In \cite{leung2012intelligent}, it aggregates relevance feedback from multiple users; while this study is useful in developing a versatile architecture, unlike the present study, it does not provide a mechanism to ensure that the search space is fully explored. In \cite{cobarzan2017interactive} it highlights the importance of humans in the search process, and acknowledges that image and video retrieval should place greater emphasis on the humans users. Our system thus further develops these approaches and provides a mechanism whereby the system learns from the humans in executing the searches.  
	
	\section{Problem Formulation}
	In a search engine system, a query needs to be input by a user via the frontend interface, through which the interaction between the system and the user triggers the dynamic involvement of the system. In general, a specific query can be submitted multiple times by the same user or by different users, where the results returned are dynamically adjusted by the system to reflect both the interests of the users’ as well as those of the system. There is a trade-off between the short-term gain obtained for the present query, and the long-term performance that will also benefit later queries. We shall employ the $\epsilon$-greedy algorithm, widely used in reinforcement learning \cite{sutton2018reinforcement}, to introduce a balance between exploitation of current knowledge to find objects that are adequately relevant, and exploration of the search space to identify the objects that are most relevant. By sacrificing a certain amount of exploitation advantage and the dwelling on a local maximum, the exploration opens the possibility of the attainment of a global maximum.
	
	Consider a given query \textit{Q} input to a multimedia search engine. We would like to examine how the dynamics of a search engine affect the returning results of the specific query \textit{Q} along with time.  We assume that there are \textit{N} multimedia objects in total in the search space. In response to the query \textit{Q}, a results list consisting of \textit{M}  objects is returned by the system, which we shall call the \textit{M}-list. We assume $\textit{N} >> \textit{M}$; i.e., a returned list is only a very small subset of the search space. We assume that the relevance of multimedia objects in the search space in relation to a given query \textit{Q} is signified by a number in the continuous scale [0, 1], and for exploitation purposes, objects having a relevance value exceeding a given threshold \textit{h} are included in the \textit{M}-list, where typical values for \textit{h} are 0.8 or 0.9. Objects that are considered to be not relevant to the query \textit{Q} typically would have a relevance value well below \textit{h}, and these objects will not be included in the \textit{M}-list through exploitation.
	
	In the $\epsilon$-greedy method, a proportion $\epsilon$ of the \textit{M}-list is used for exploration, while ($1 - \epsilon$) is used for exploitation. Here, we assume that the ordering of objects within the \textit{M}-list is not important; this is particularly true for image objects whereby users tend not to just go through the first few objects on the list, but also the remaining objects as well. Repeated presentations of the \textit{M}-list in response to the query \textit{Q} (possibly from different users) are denoted by the $\textit{M}_1, \textit{M}_2, \textit{M}_3$, ..., where $\textit{M}_i$ signifies the \textit{i}th \textit{M}-list presented for the query \textit{Q}.  We let \textit{r} = $\epsilon$\textit{M}, and \textit{K} = ($1 - \epsilon$)\textit{M}, i.e, \textit{M} = \textit{r} + \textit{K}, and we include \textit{r} randomly chosen objects in the \textit{M}-list, where each available object apart from the \textit{K} objects from exploitation is chosen with equal probability. 
	
	We consider two variations of the $\epsilon$-greedy exploration algorithm.
	\begin{enumerate}[label=\Alph*]
		\item Exploration with object re-selection. Each presentation of the \textit{M}-list with the \textit{r} random objects are done in such a way that when a given random object \textit{Z} has been included in a previous \textit{M}-list presentation, it can be re-selected for inclusion in a subsequent \textit{M}-list presentation. 
		\item Exploration without object re-selection. Each presentation of the \textit{M}-list with the \textit{r} random objects are done in such a way that when a given random object \textit{Z} has been included in a previous \textit{M}-list presentation, it will be excluded for inclusion in a subsequent \textit{M}-list presentation.
	\end{enumerate}
	
	The details of these are given respectively in algorithms EGSE-A, and EGSE-B ($\epsilon$-greedy algorithm for search exploration). Compared with EGSE-B, EGSE-A has a greater degree of fault-tolerance in that if a user inadvertently overlooks a highly relevant random object, it can still be discovered in a subsequent presentation. On the other hand, such duplication of effort will tend to slow down search space exploration. Consequently, EGSE-B has the advantage that the exploration can advance at a faster pace but supports less fault-tolerance since if the user somehow overlooks the most relevant object, it will remain hidden and never be discovered.
	
	More precisely, let \textit{X} be the multimedia object that is most relevant to the query \textit{Q}, but the relevance value of \textit{X} is currently well below the threshold value \textit{h}. Under the $\epsilon$-greedy algorithm, we seek to provide answers to the following questions:
	\begin{itemize}
		\item What is the probability that \textit{X} can be first discovered on the \textit{k}th presentation of the \textit{M}-list? That is, we would like to find the probability
		\begin{displaymath}
		\mathbb{P}\{X \in M_k: X \notin M_1, ..., X \notin M_{k-1} \}
		\end{displaymath}
		\item What is the average time for \textit{X} to be discovered (i.e. the average time for \textit{X} to be included on an \textit{M}-list for the first time)?  That is, we wish to find \textit{k} such that
		\begin{displaymath}
		\min_{k}\{M_k: X \in M_k \}.
		\end{displaymath}
	\end{itemize}
	In the next section, we shall derive solutions to these measures.
	
	\section{Probabilistic Analysis}
	\subsection{Performance Analysis of Algorithm EGSE-A}
	Consider $\epsilon$-greedy Algorithm EGSE-A. The probability that \textit{X} is included in a particular \textit{M}-list is given by:
	\begin{displaymath}
	a_{r,M} = \frac{{N-M+r-1\choose r-1}}{{N-M+r\choose r}},
	\end{displaymath}
	which is obtained as follows. The number of objects in the pool of objects from which the \textit{r} random objects are selected is $N-(M-r)=N-M+r$. The total number of combinations in choosing \textit{r} objects from $(N - M + r)$ objects including \textit{X} is ${N-M+r-1\choose r-1}$, since we are excluding \textit{X} from the pool of selections, and then always including \textit{X} in the resultant chosen \textit{r} objects by selecting only ($r-1$) remaining objects and reserving a place for \textit{X}. This is then divided by the total number of possibilities ${N-M+r\choose r}$ in choosing any \textit{r} objects from among the $(N - M + r)$ objects. Letting $\beta_{r,M}$ be such that $\alpha_{r,M}$ + $\beta_{r,M}$ = 1, then denoting by $\textbf{\textit{U}}_{r,M}$ the random variable signifying the time to discover \textit{X} (for the first time), we have 
	\begin{equation}
	\mathbb{P}[\textbf{U}_{r,M} = k] = \alpha_{r,M}\beta^{k-1}_{r,M},
	\end{equation}
	with corresponding probability generating function
	\begin{displaymath}
	F(z) = \frac{\alpha_{r,M}z}{1-\beta_{r,M}}.
	\end{displaymath}
	
	The mean and variance of $\textbf{\textit{U}}_{r,M}$ can thus be obtained by differentiation \cite{feller2008introduction}:
	\begin{equation}
	\mathbb{E}[\textbf{U}_{r,M}] = \frac{{N-M+r\choose r}}{{N-M+r-1\choose r-1}},
	\end{equation}
	\begin{align}
	\mathbb{V}[\textbf{U}_{r,M}] &= \left[\frac{{N-M+r\choose r}}{{N-M+r-1\choose r-1}}\right]^2 \nonumber
	\\  &\times\bigg[\frac{{N-M+r\choose r} - {N-M+r-1\choose r-1}}{{N-M+r\choose r}}\bigg].
	\end{align}  
	
	\begin{algorithm}
		\caption{\textbf{EGSE-A}: Search Space Exploration with Constant Probability}
		\begin{algorithmic}[1]
			\Require epsilon E, length of result list \textit{M}, query max counter \textit{C}
			\State Initialize terminating condition $\Delta$ $\gets$ \textit{False}, 
			\State Initialize query counter $\Theta$ $\gets$ 0,  
			\State Initialize exploration proportion $R \gets E \times M$,
			\State Initialize exploitation proportion $K \gets (1-E) \times M$,
			\While{$\Delta$ == \textbf{False}	}
			\State Retrieve and parse new user query \textit{Q}
			\State Determine $S_1$ = \{$O_i$ $\vert$ objects with the highest relevant scores$\}_{i=1}^k$, where $\vert S_1 \vert = K$ 
			\State Determine $S_2$ = \{$O_j$ $\vert$ $O_j \in S_1^\complement\}_{j=1}^R$, where $\vert S_2 \vert = R$
			\State Present $M$-list := $S_1 \cup S_2$ to user
			\State Capture object click information from user
			\State Increment the score of clicked objects
			\State $\Theta$ $\gets$ $\Theta + 1$
			\If{$\Theta ==$ {\textit{C}}}
			\State $\Delta \gets$ \textit{True}
			\EndIf
			\EndWhile		
		\end{algorithmic}
	\end{algorithm}

	The probability that \textit{X} is discovered in finite time can likewise be obtained directly from the probability generating function and is found to equal to one.
	
	\subsection{Performance Analysis of Algorithm EGSE-B}
	To analyze EGSE-B, we shall determine the probability of \textit{X} being included in an \textit{M}-list for the first time, bearing in mind that objects through exploration which have been presented in earlier \textit{M}-lists are marked and excluded from further presentation. We denote by $f_{r,M,k}$ the following first passage probability
	\begin{displaymath}
	f_{r,M,k} = \mathbb{P}\{X \in M_k: X \notin M_1, ..., X \notin M_{k-1} \}
	\end{displaymath}
	For example, we have
	\begin{align}
	f_{r,M,3} &= \mathbb{P}\{X \in M_k: X \notin M_1, X \notin M_{2} \} \nonumber
	\\ &= \frac{{N-M+r-1\choose r}}{{N-M+r\choose r}} \times \frac{{N-M-1\choose r}}{{N-M\choose r}}
	\times \frac{{N-M-r-1\choose r-1}}{{N-M-r\choose r}} \nonumber
	\end{align}
	
	The above is obtained as follows. The first factor represents the probability of $X \notin M_1$: the numerator is the number of combinations of all object choices from $N - M + r$ -1 objects, excluding $X$, from which the system selects $r$ objects; the denominator represents the unrestricted choice of $N - M + r$ objects from which the system selects $r$; this factor represents the probability of not including $X$ in the first $M$-list. The second factor represents the probability of $X \notin M_2$: the numerator is the number of combinations all object choices from the now $N-M-1$ objects, since the first $r$ objects presented in $M_1$, together with $X$ are excluded, and from which we select $r$ objects;  the denominator represents the unrestricted choice of $N - M$ objects from which the system selects $r$, where the first $r$ objects has been excluded; this factor represents the probability of not including $X$ in the second $M$-list. The third factor represents the probability of $X \in M_3$: the numerator is the number of combinations all object choice from the now $N - M - r - 1$ objects, since the first 2$r$  objects presented in $M_1$ and $M_2$, together with $X$ are excluded, and from which we select $r-1$ objects, since a place is now reserved for $X$; the denominator represents the unrestricted choice of $N - M - r$ objects from which the system selects $r$, where the first 2$r$ objects have been excluded; this factor represents the probability of successfully including $X$ in the third $M$-list for the first time. Using the above reasoning, we can establish a recurrence relation for $f_{r,M,k}$:
	\begin{align}
	f_{r,M,k+1} &= f_{r,M,k} \nonumber \times \frac{{N-K-kr\choose r}}{{N-K-kr-1\choose r-1}}
	\\ &\times \frac{{N-K-kr-1\choose r}}{{N-K-kr\choose r}}
	\times \frac{{N-K-(k+1)r-1\choose r-1}}{{N-K-(k+1)r\choose r}}, 
	\end{align}
	where the second factor serves to remove the successful inclusion probability in $f_{r,M,k}$, and then replace this success probability by a failure to include $X$ probability, which is the third factor. The final factor gives the successful inclusion probability at the $(k+1)$th presentation after $k$ failed attempts to include $X$ before.
	
	The solution to the above recurrence relation requires rather involved manipulations and is relegated to Appendix A, which also provides the derivation for the mean and variance of $\textbf{V}_{r,M}$, the random variable signifying the time to discover $X$ (for the first time) under EGSE-B,  
	
	\begin{algorithm}
		\caption{\textbf{EGSE-B}: Search Space Exploration with Variable Probability}
		\begin{algorithmic}[1]
			\Require epsilon E, length of result list \textit{M}, query max counter \textit{C}
			\State Initialize terminating condition $\Delta$ $\gets$ \textit{False}, 
			\State Initialize query counter $\Theta$ $\gets$ 0,  
			\State Initialize exploration proportion $R \gets E \times M$,
			\State Initialize exploitation proportion $K \gets (1-E) \times M$,
			\State Initialize previously presented $M$-list for Query $Q_i$ as $S_i \gets \emptyset$, for all possible $i$
			\While{$\Delta$ == \textbf{False}	}
			\State Retrieve and parse new user query $Q_i$
			\State Determine $S_1$ = \{$O_l$ $\vert$ objects with the highest relevant scores$\}_{l=1}^k$, where $\vert S_1 \vert = K$ 
			\If {$\vert(S_1 \cup S_i)^\complement\vert \geq R$}
			\State Determine $S_2$ = \{$O_j$ $\vert$ $O_j \in (S_1 \cup S_i)^\complement\}_{j=1}^R$, where $\vert S_2 \vert = R$
			\Else
			\State Determine $S_2 =(S_1 \cup S_i)^\complement$, where $\vert S_2 \vert = \vert(S_1 \cup S_i)^\complement\vert$
			\EndIf
			\State Present $M$-list := $S_1 \cup S_2$ for query $Q_i$ to user
			\State $S_i \gets S_i \cup S_1 \cup S_2$
			\State Capture object click information from user
			\State Increment the score of clicked objects
			\State $\Theta$ $\gets$ $\Theta + 1$
			\If{$\Theta ==$ {\textit{C}} or $(S_1 \cup S_i)^\complement == \emptyset$}
			\State $\Delta \gets$ \textit{True}
			\EndIf
			\EndWhile		
		\end{algorithmic}
	\end{algorithm}
	
	\begin{equation}
	\mathbb{E}[\textbf{V}_{r,M}] = \frac{N-M+2r}{2r},
	\end{equation}
	\begin{equation}
	\mathbb{V}[\textbf{V}_{r,M}] = \frac{1}{12}\times\biggl\{\bigg[\frac{N-M+r}{r}\bigg]^2-1\biggr\}.
	\end{equation}
	Since EGSE-B explores the complete search space faster than that of EGSE-A, and since EGSE-A exhausts the search space in finite time with probability one, EGSE-B will also exhaust the search space in finite time with probability one. 
	
	\section{Experiments and Results Interpretation}
	In this section, we describe both the experimental work performed on real dataset and simulations to evaluate the effectiveness and performance of the above algorithms. While real dataset experiments stress on the worst case scenario analysis, simulations aim to verify the theoretical results on the average performance.
	
	\subsection{Real dataset Experiments}
	\noindent \textbf{Dataset and Preprocessing.} To build up a real dataset suitable for index-based multimedia search, we crawl both Google and Bing to retrieve multimedia objects under the same motif with descriptive information. With the crawled raw data, we form a musical theme based image dataset by carefully selecting relevant images along with their true labels. Specifically, each image falls into one of four categories: $\{$\textit{"grand piano", "upright piano", "classical guitar", "harp"}$\}$, where each category takes up 25$\%$ of the whole dataset. Such images are associated together in many semantic contexts and often appear together in searches such as "\textit{Le quattro stagioni} orchestra". To ensure the training quality and to deal with the significantly varied image sizes, all images are re-scaled to $512 \times 512$. Fig. 1 shows sample images of the final dataset. 
	
	\vskip 0.1cm \noindent \textbf{Experimental Details.} To randomize the initialization of relevance index values (RIVs) for all possible semantic index terms, a Gaussian distribution is utilized. In order to study the effectiveness of exploration using the proposed $\epsilon$-greedy algorithms, we are interested in the case whether a hidden object with unfavorable indexes can be successfully discovered. Therefore, we randomly select one object with the true label that will be included in the input query and artificially change its label to be a misleading one. To utilize the associated textual information, RIVs for images with the target true label are increased with a calibrated delta value within one standard deviation of the Gaussian distribution. \textit{Min-max normalization} is also adopted so that RIVs are normally distributed within the [0,1] interval:
	\begin{displaymath}
	v = \frac{v - min(v)}{max(v) - min(v)}.
	\end{displaymath}	
	The distribution of initial RIVs are presented in Fig. 3.
	
	The process starts by inputting a query containing one of the above four categorical labels. As shown in previous sections, returned lists are comprised of a greedy part for exploitation and an exploratory portion defined by $\epsilon$. Each time when a \textit{M}-list is returned, evaluative information from users is provided to allow successful evolution of indexes. Here, we assume that users provide implicit feedback for the greedy part only if they find objects that are of interest, and explicit feedback would be directly given for evaluating objects from the exploratory portion. 
	
	Figures 4 and 5 demonstrate the evolution of the above process, where the input queries contain the keyword \textit{"grand piano"}. Fig. 4 shows that initially, RIVs for each category are nearly uniformly distributed, showing no bias towards any of them. To represent the user evaluations, we randomly choose zero to five objects in the greedy part representing the clicking behavior, either boosting their RIVs with a small delta value, if their true labels are \textit{"grand piano"}, or decreasing the RIVs otherwise as punishment. The same procedures apply to the $\epsilon$ explorative portion, except that all objects from this portion are evaluated explicitly. In particular, the parameters are set as $N=1000, M=50, \epsilon=0.1$. We are interested in the analysis of the performance of EGSE-A and EGSE-B under the worst case scenario, which can be measured by the number of queries required to discover the target hidden object $X$. By worst case scenario, it is meant that:
	\begin{itemize}
		\item $X$ never appears in the greedy exploitation component of the \textit{M}-list,
		\item $X$ requires the maximum number of queries to be discovered under exploration.
	\end{itemize}
	
	With the above parameters, under EGSE-B, 200 queries are needed for $X$ to be presented with probability one. In practice, $X$ can be successfully discovered with less than 200 queries in most cases due to the relaxation of the above two conditions. To corroborate the claim, we choose several random seeds to run the experiments for EGSE-A and EGSE-B under the same set of parameters, and report the result for the one that requires most quries for EGSE-B. With the same random seed, EGSE-B runs 172 queries to discover the object $X$, whereas when object re-selection is permitted in EGSE-A, the number of query times increases to 223 to discover $X$. We see that X can still be discovered within a reasonable number of queries with EGSE-A. The increase can be regarded as the price paid for supporting greater fault-tolerance for the EGSE-A algorithm. 
	
	At the time when $X$ is discovered, Fig. 5 shows the distribution of RIVs for each category. The fact that the overall value scale for \textit{"grand piano"} is significantly higher than other categories suggests during the evolution process, the proposed EGSE-A and EGSE-B algorithms not only allow successful discovery of object $X$, but also successfully separate out the indexes of true interest from the irrelevant ones compared to the initial uniform pattern across different categories. Notably, the initially unrealistically high RIVs for irrelevant labels are flattened in the end, confirming the efficacious balance between exploration and exploitation.
	
	Meanwhile, it is also interesting to find that the evolution of query precision for EGSE-A and EGSE-B manifests the same pattern, as shown in Fig. 6. Generally, the precision for the index-based multimedia search climbs up from extremely low levels to much more precise values, ending in providing satisfactory search results. As EGSE-B typically discovers $X$ much faster than EGSE-A, the time that the system spent on evolution starting from the initial state is not as long as that of EGSE-A. If we are only concerned with the discovery of $X$, the query precision would converge to around 82\%. On the other hand, if longer time for learning and searching development is allowed, precision can reach 92\%. The final \textit{M}-list using EGSE-A is provided in Fig. 2, where the hidden object is marked with a read label.  
	
	\begin{figure*}[h]
		\centering
		\includegraphics[width=.49\textwidth]{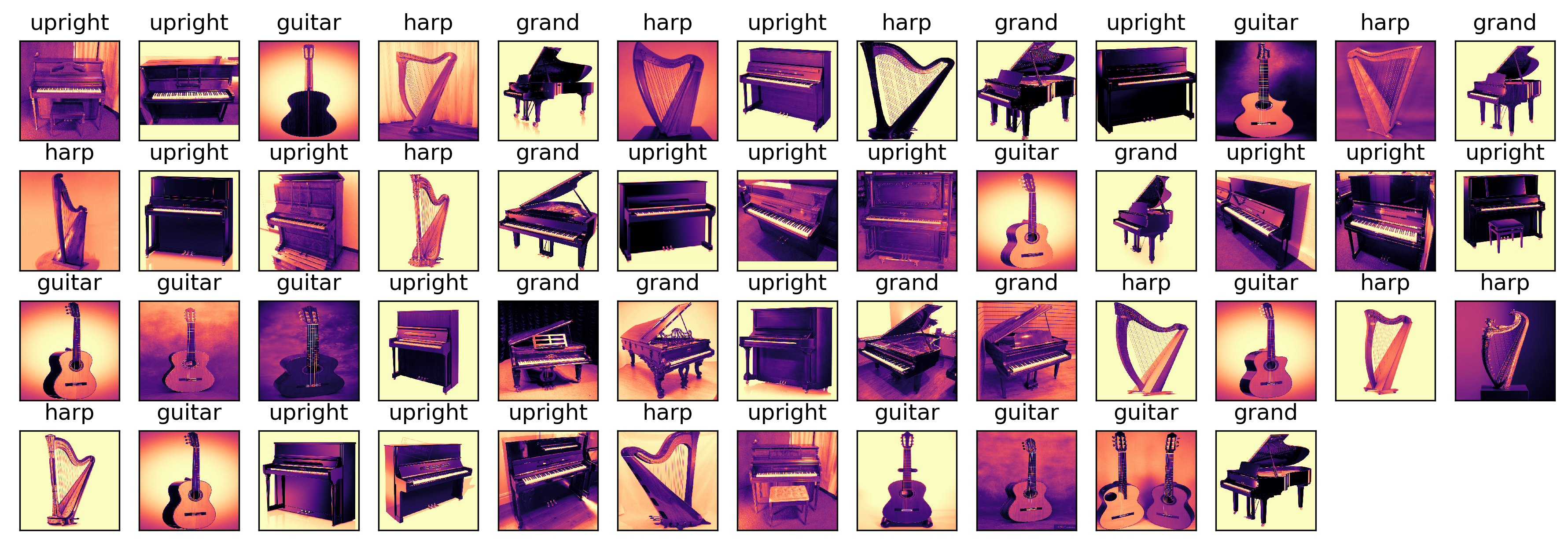}
		\includegraphics[width=.49\textwidth]{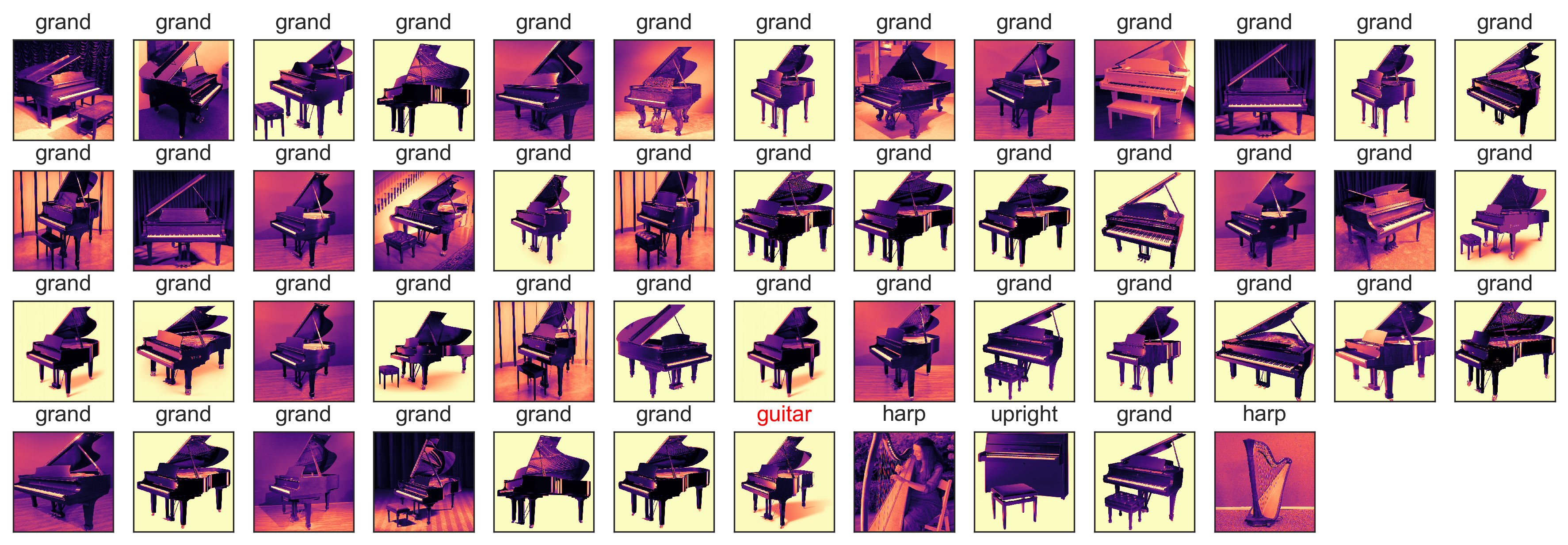}
		\caption*{Fig. 1. Sample Images from Dataset (Size = 50). Fig. 2. Final Returned \textit{M}-list using EGSE-A. Settings: $N = 1000, M = 50, \epsilon = 0.1$.}
	\end{figure*}
	
	\begin{figure*}[h]
		\centering
		\includegraphics[width=.23\textwidth]{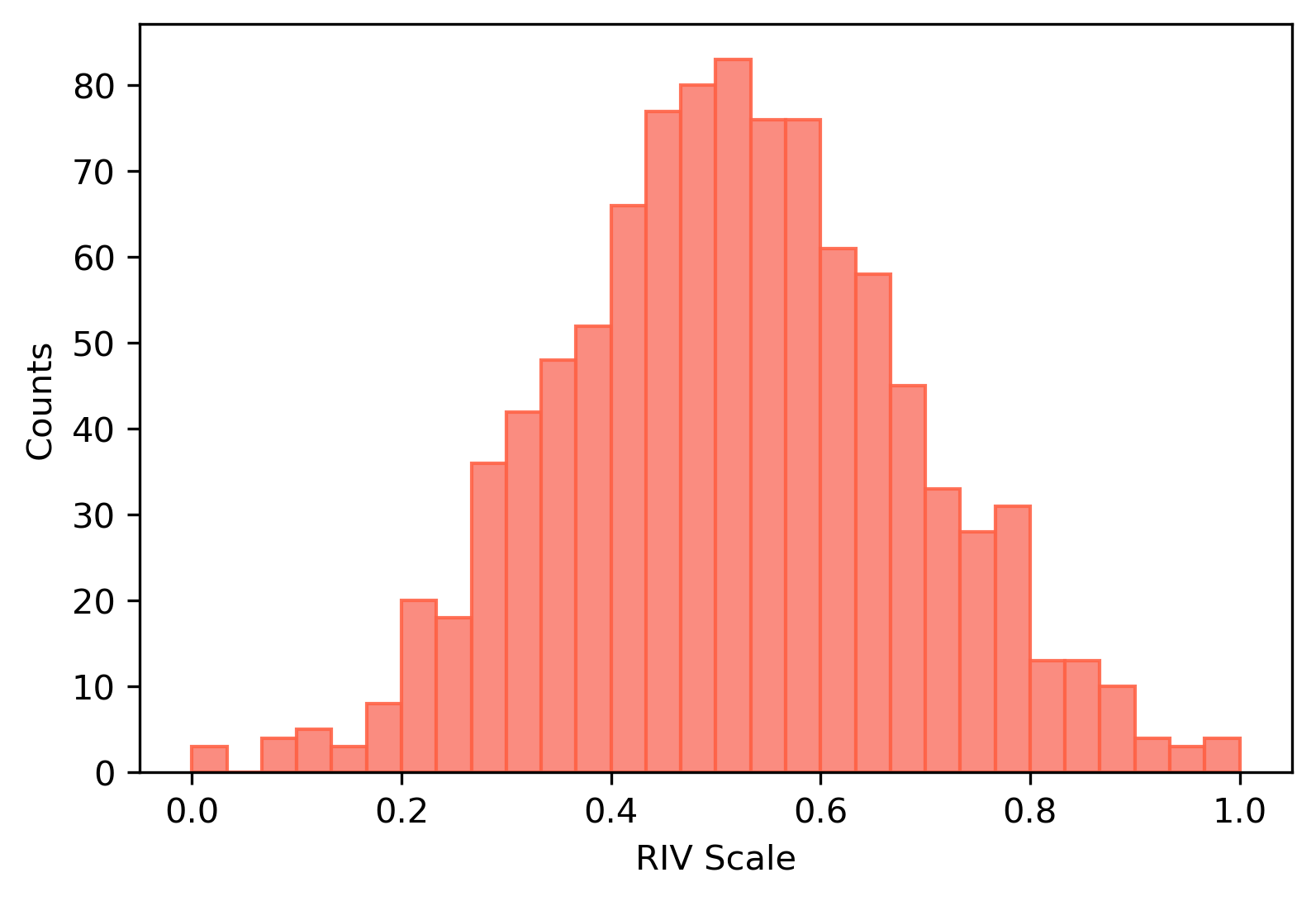}
		\includegraphics[width=.23\textwidth]{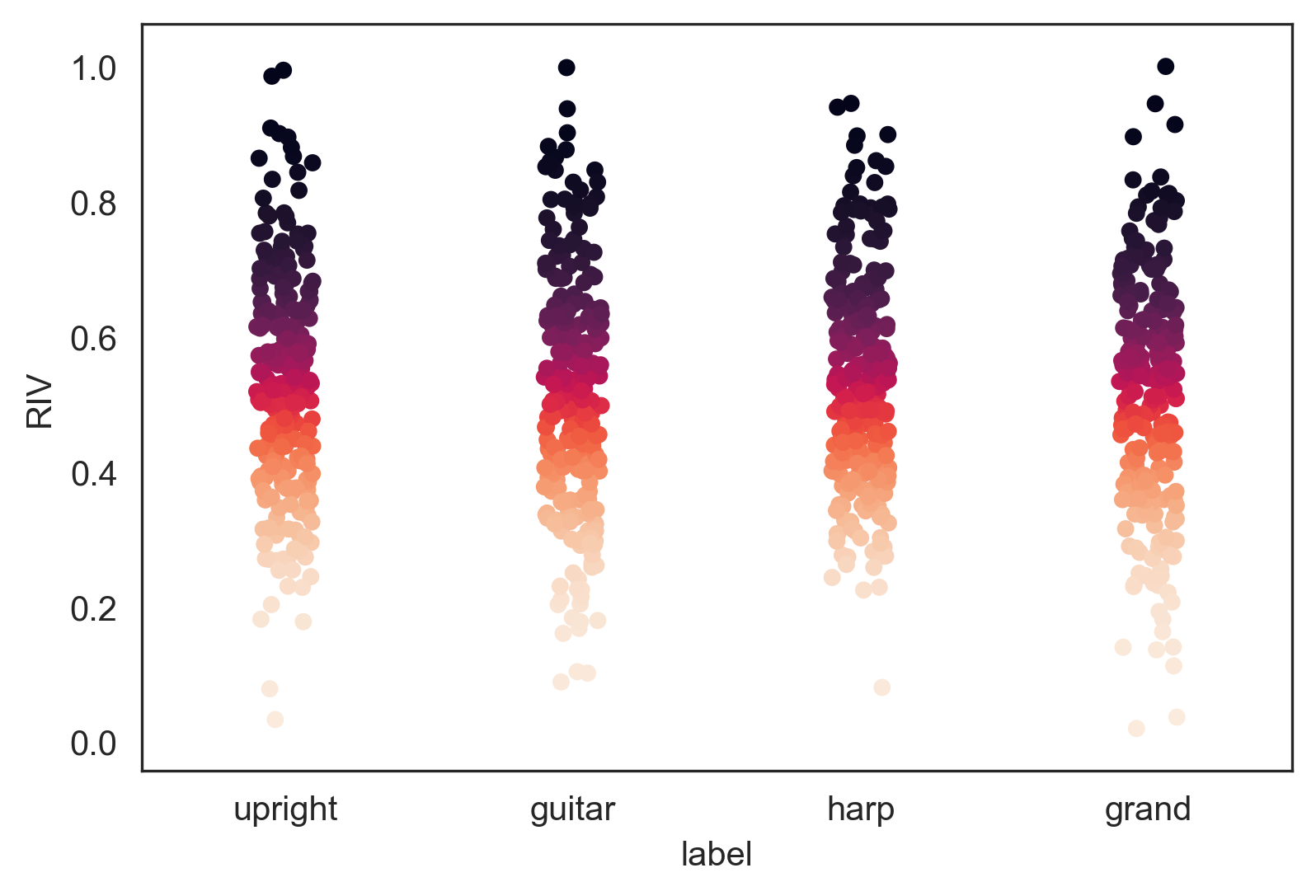}
		\includegraphics[width=.23\textwidth]{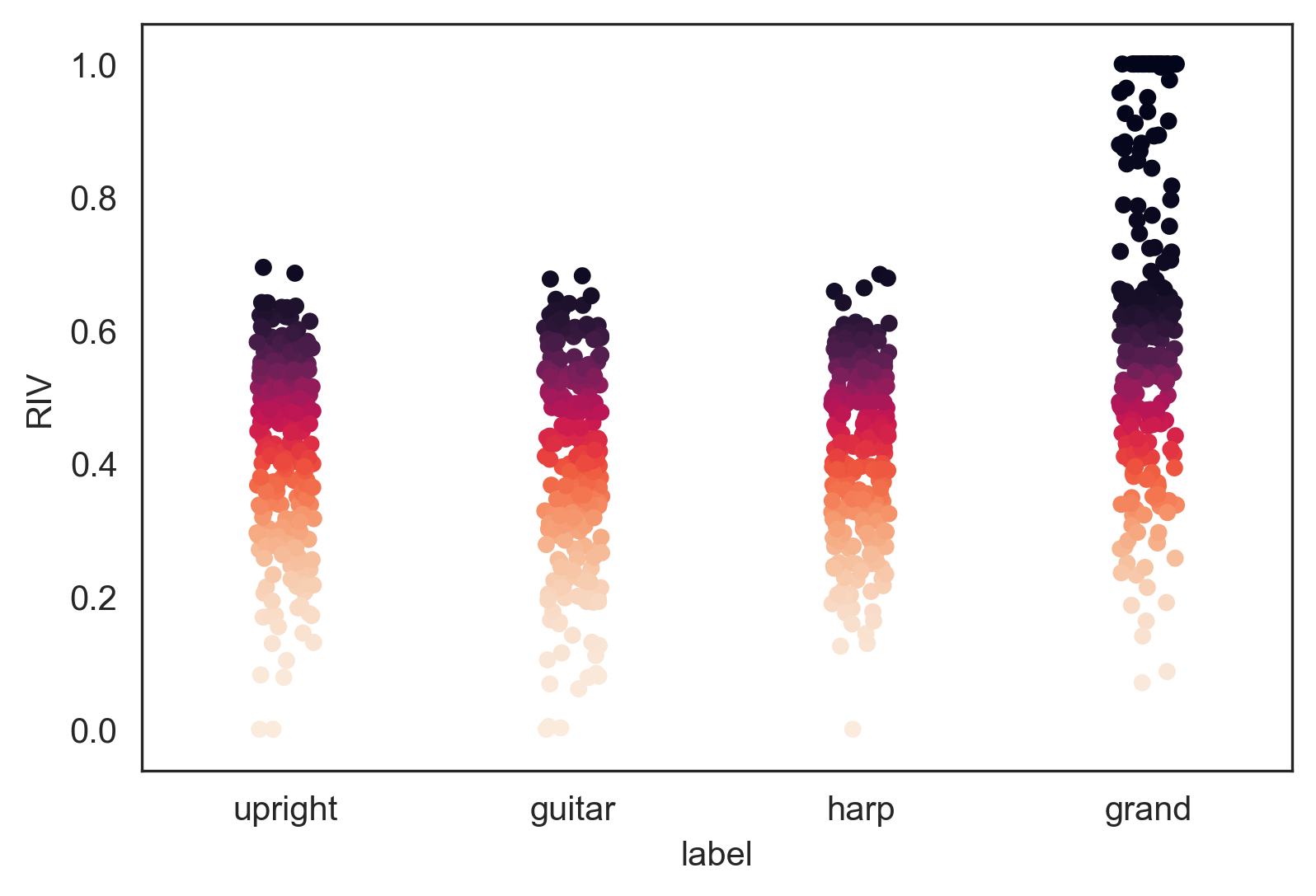}
		\includegraphics[width=.23\textwidth]{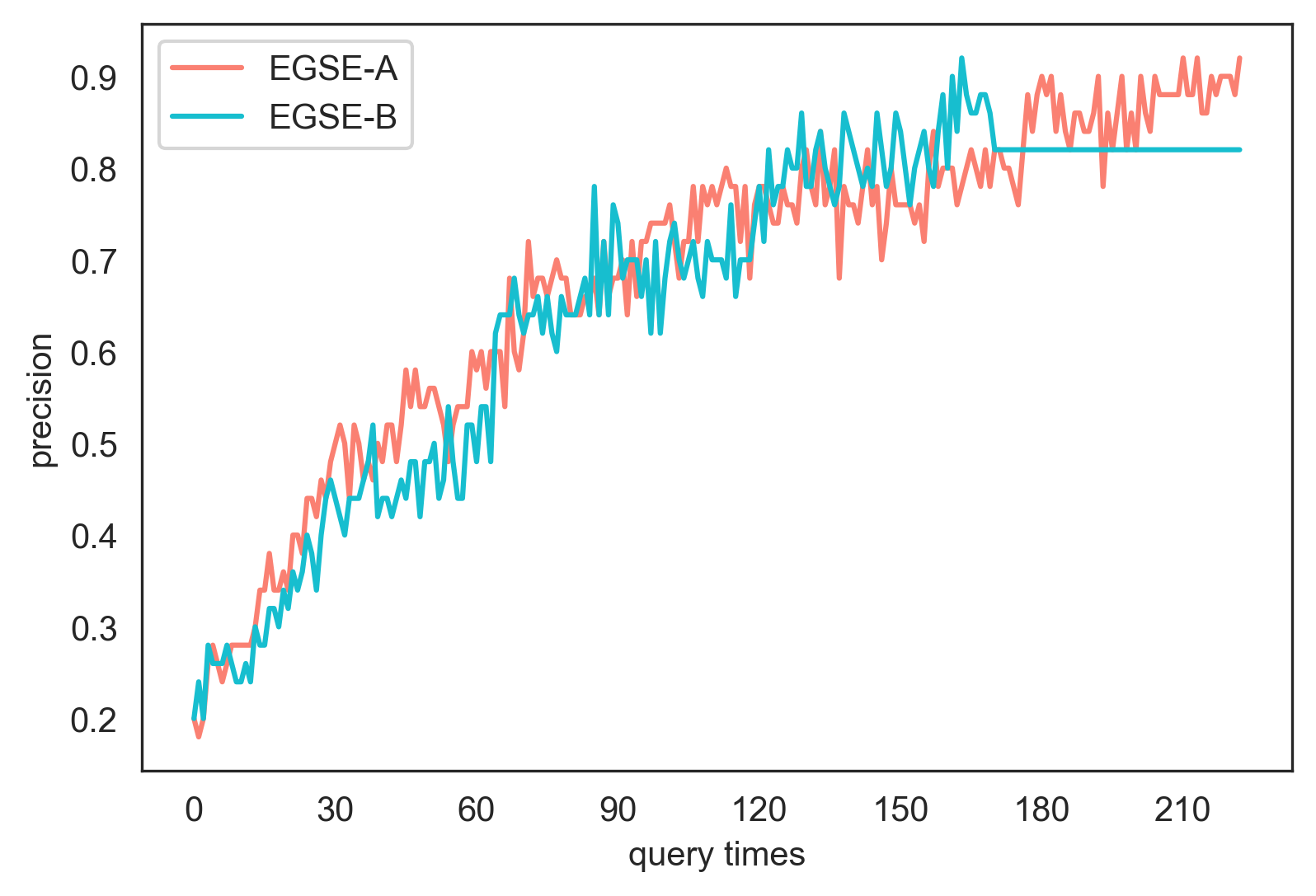}
		\caption*{Fig. 3. Distribution of Initial RIV Scores. Fig. 4. Distribution of Initial RIV Scores for Each Category (EGSE-B). Fig. 5. Distribution of RIV Scores for Each Category When Hidden Object $X$ is Discovered (EGSE-B). Fig. 6. Evolution of Query Precisions against Query Times. Settings: $N = 1000, M = 50, \epsilon = 0.1$.}
	\end{figure*}
	
	\begin{figure*}[h]
		\centering
		\includegraphics[width=.23\textwidth]{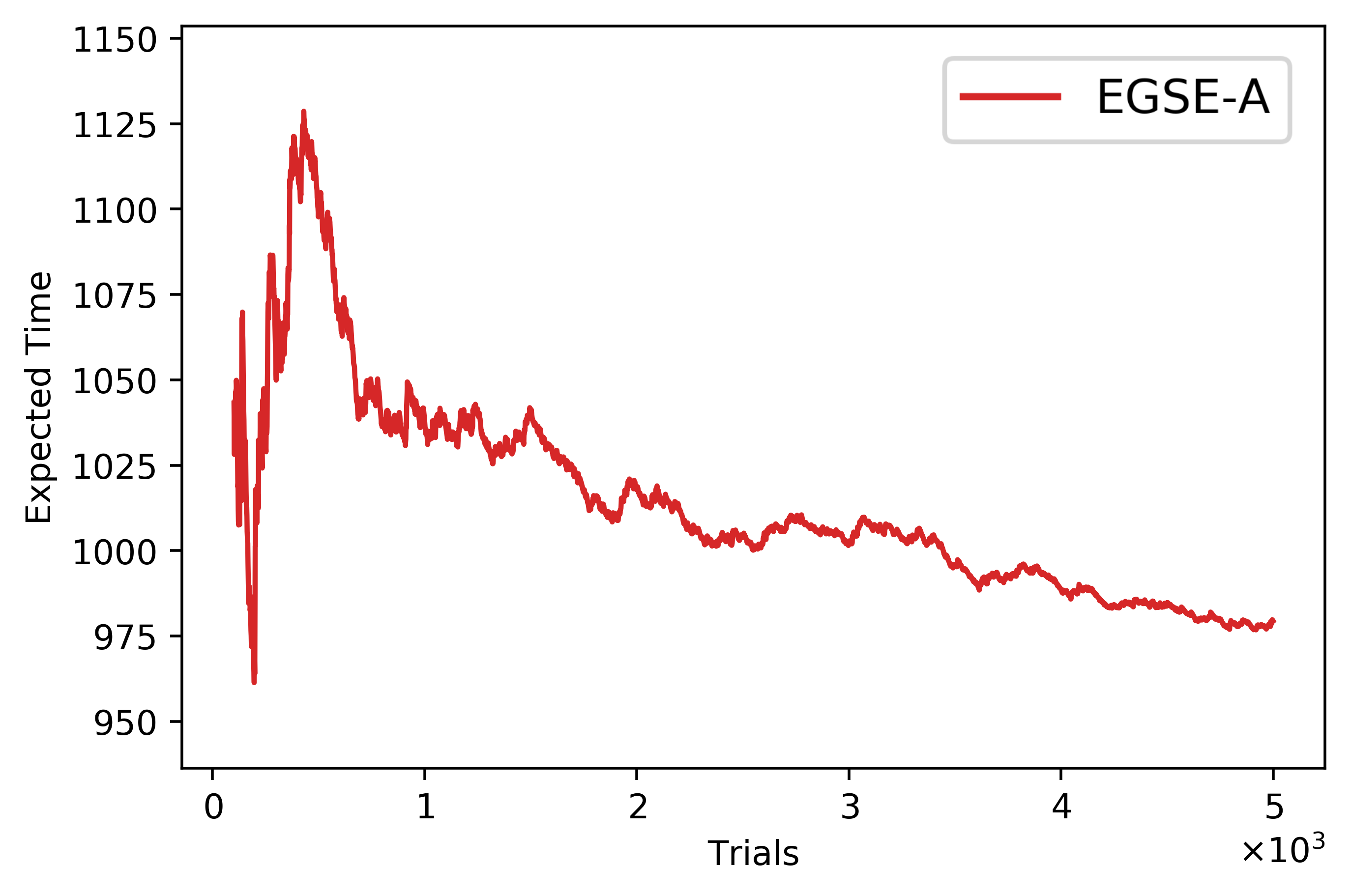}
		\includegraphics[width=.23\textwidth]{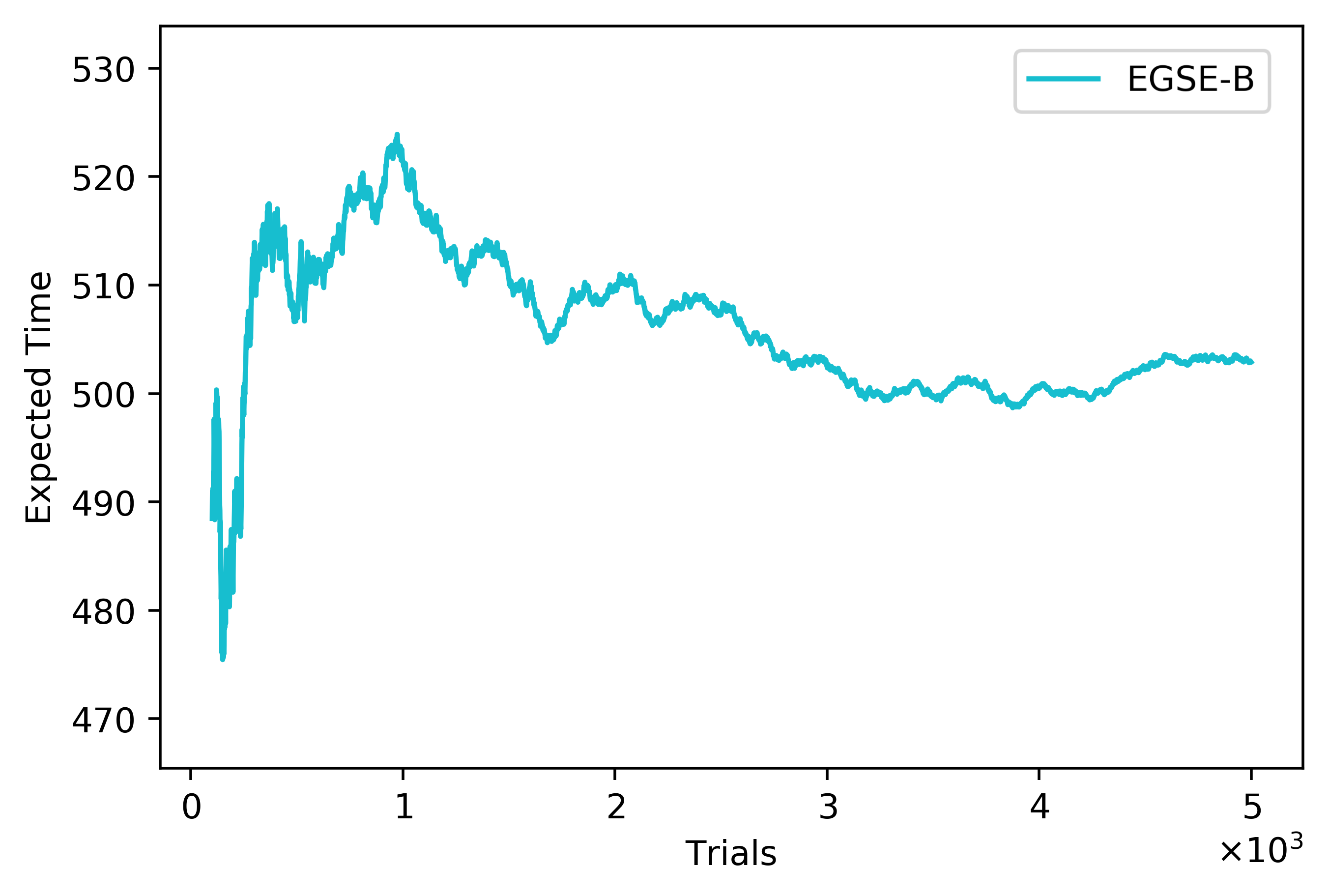}
		\includegraphics[width=.23\textwidth]{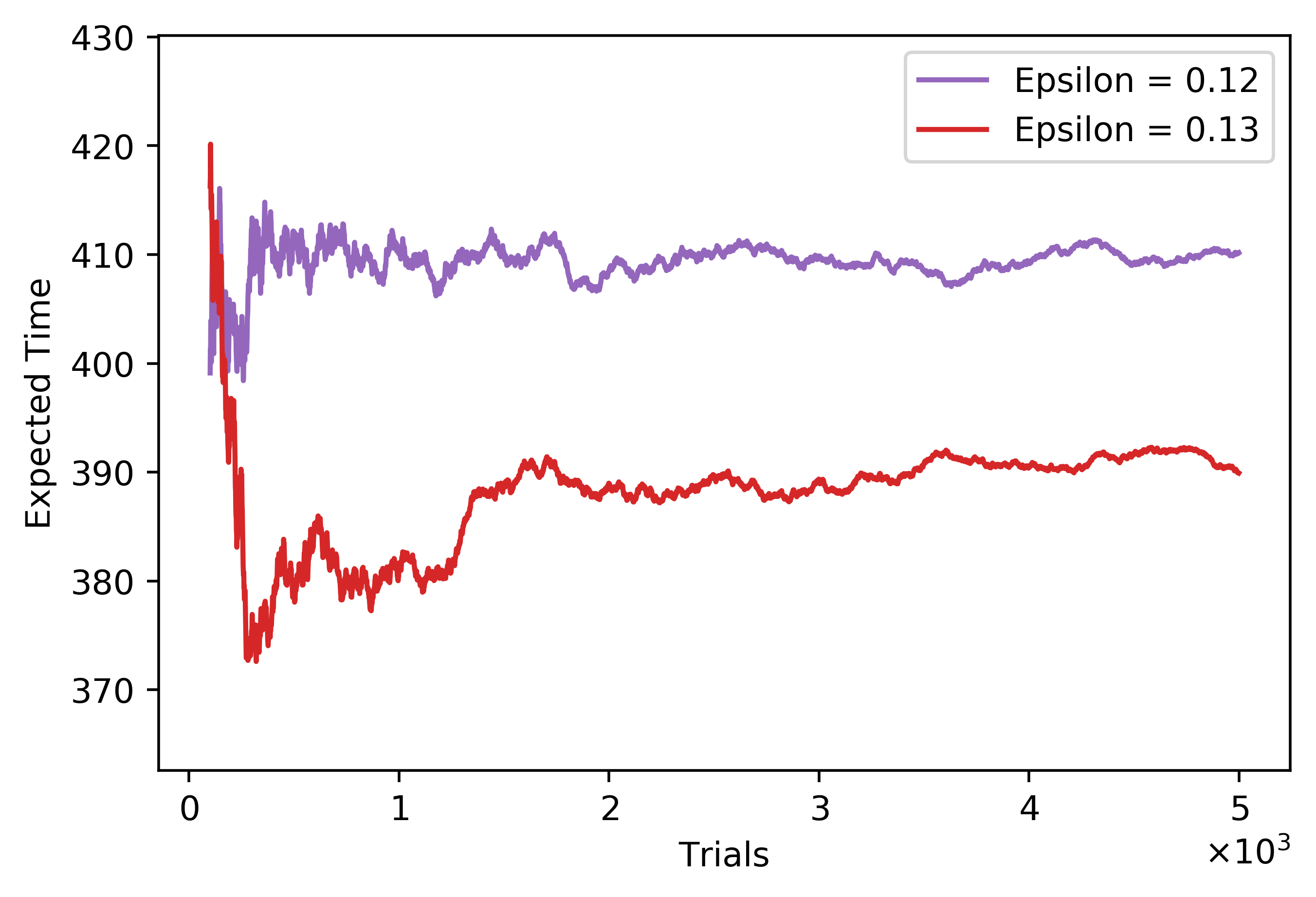}
		\includegraphics[width=.23\textwidth]{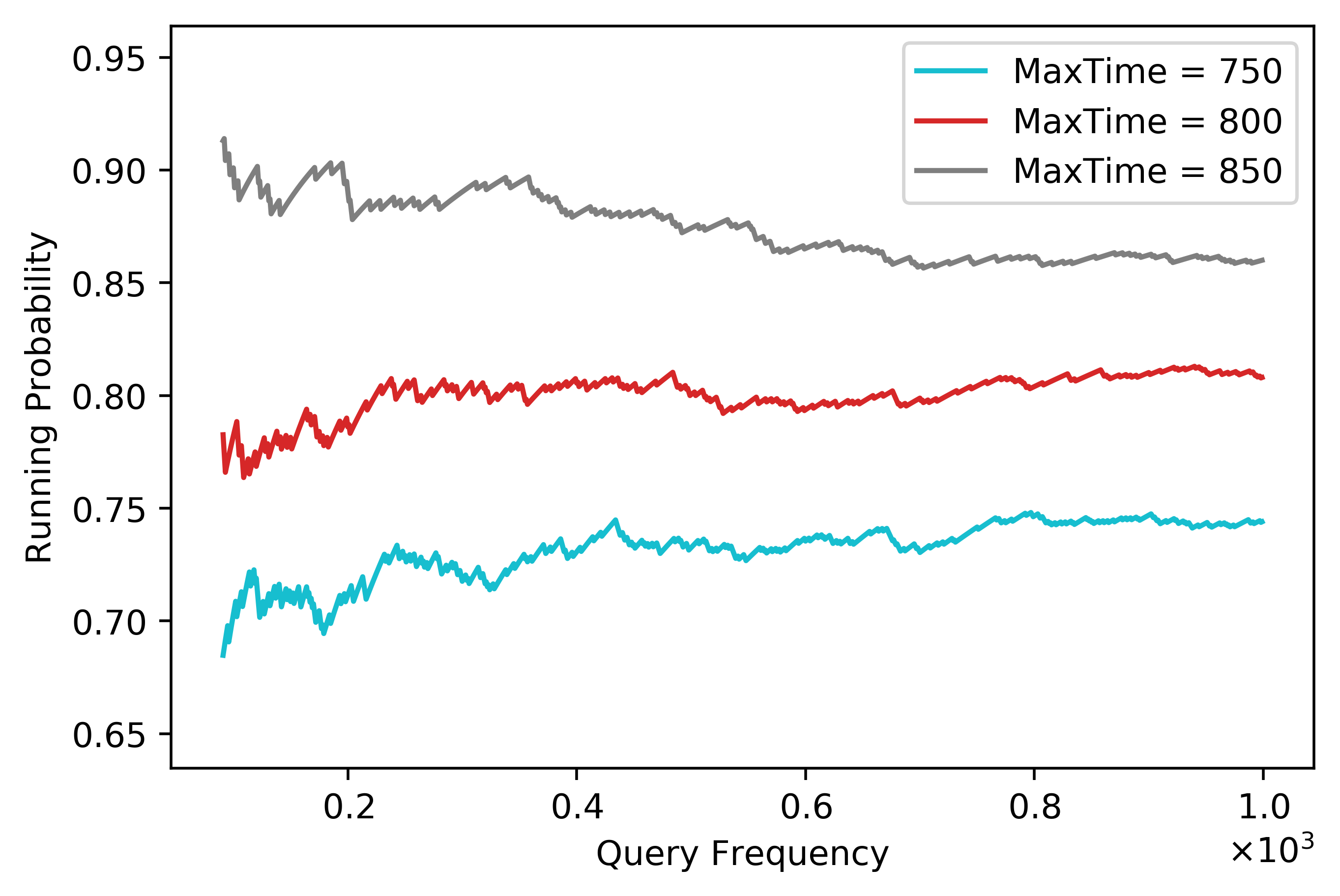}
		\caption*{Fig. 7. Expected Discovery Time of EGSE-A. Fig. 8. Expected Discovery Time of EGSE-B. Settings: N = 10000, M = 100, $\epsilon$ = 0.1. Fig. 9. Expected Discovery Time of EGSE-B with $\epsilon$ = \{0.12, 0.13\}. Fig. 10. Probability of Discovering the Most Relevant Object in EGSE-B with Time Constraints.}
	\end{figure*}
	
	\subsection{Monte-Carlo Simulations}
	\noindent\textbf{Experimental Setup.} To evaluate how EGSE-A and EGSE-B perform in environments with a large degree of stochastic influence, we adopt the Monte-Carlo method to efficiently generate samples for simulating the formation of query results. In particular, four different simulations that examine different aspects of the problem are presented:
	\begin{itemize}
		\item Case I: Evaluate $\mathbb{E}[\textbf{U}_{r,M}$] for EGSE-A.
		\item Case II: Evaluate $\mathbb{E}[\textbf{V}_{r,M}$] for EGSE-B.
		\item Case III: Effect of varying $\epsilon$ for E[$\textbf{V}_{r,M}$] of EGSE-B.
		\item Case IV: Probability of discovering $X$ with constraints.
	\end{itemize}
	To ensure that the behavior is in accordance with the corresponding real-world scenario in the long run, each testing scenario is simulated with 5,000 trials for Cases I to III, and 1,000 trials for Case IV, where the the convergence behavior can be observed. Because of the presence of randomness, the discovery time of the interested object $X$ can vary even with the same settings of parameters. Examining the expected discovery time can reveal the general effectiveness of the strategies. Meanwhile, $\epsilon$ is varied to check whether the strategy involved is able to converge effectively under various circumstances. During the experiments, we find that EGSE-A and EGSE-B with different values of $\epsilon$ exhibit somewhat similar patterns, hence for brevity we report only the results of EGSE-B. Meanwhile, time constraints are set to evaluate whether the discovery of $X$ can be finished in finite time steps.
	
	\vskip 0.1cm \noindent\textbf{Results and Interpretations.} The expected discovery time of EGSE-A and EGSE-B under the same parameter settings are shown in Fig. 7 and Fig. 8. Here, we let \textit{N} = 10,000, \textit{M} = 100 and $\epsilon$ = 0.1. The evolution of the expected discovery time is plotted against the number of trials. In the inception of both EGSE-A and EGSE-B, the expected discovery time fluctuates tremendously and the results can be seen to vary dramatically with the relatively small number of trials. With the number of trials increasing, the patterns of EGSE-A and EGSE-B gradually become steady and finally converge to the theoretical values of 991.0 (with relative error 0.036\%), and 496.0 (with relative error 1.428\%) respectively. These results suggest that when the same query is visited a sufficient number of times, both EGSE-A and EGSE-B are guaranteed to return the most relevant object in the search space regardless of the initial settings. With the same set of parameters, EGSE-B takes a much smaller discovery time to discover the most relevant object compared to EGSE-A.
	
	To see how the discovery time evolves with different $\epsilon$ values, Fig. 9 shows the expected discovery time of EGSE-B with $\epsilon$ values of 0.12 and 0.13. Again, fluctuations exist only in the early stage when the specific query is input for relatively few times. With the increasing number of trials, the expected discovery times are around 410.17 and 389.92 with relative errors  of 0.804\% and 0.655\% respectively compared to the theoretical values of 413.5 and 381.77. The same evolution pattern holds true for EGSE-A as well. As a result, no matter which variant of the $\epsilon$-greedy method is adopted, the value choice of $\epsilon$ would only affect how aggressive the exploration strategy is but has no impact on uncovering the most relevant object in the end.
	
	In real-world multimedia search systems, it is often desirable to return the most satisfactory results within limited time and resources, so that users would deem it as effective. Therefore, it is of interest to know whether object $X$ can be discovered within limited time steps. Specifically, we use the same settings as for Fig. 7 and Fig. 8, and set the maximum time step to be 750, 800 and 850 respectively for each query trial. The results are plotted in Fig. 10, which shows that along with time, the probabilities of discovering object $X$ under such time constraints are respectively 75.8\%, 80.5\% and 86.1\%. It agrees with the intuition that tighter constraints result in smaller probability of discovering the desired object, as the unfavorable initial settings would always require longer time for the discovery. Nevertheless, even in systems where time resources is a significant concern, the modified $\epsilon$-greedy algorithms can still lead to the promising discovery of the desired object. This in turn corroborates the effectiveness of EGSE-A and EGSE-B.

	\section{Summary and Conclusions}
	Multimedia data objects tend to possess rich and diverse attributes which make them difficult to be fully indexed. As a result, directly presenting to the user multimedia objects which seem relevant may not be optimal since the most relevant objects for given queries may escape notice and never be shown or retrieved. This will result in the landing in a local maximum, whereby suboptimal results are repeatedly shown and received an increase in relevance score, while the most relevant objects are never exposed and stand no chance of being clicked to receive an increase in relevance score. Consequently, to overcome this problem, the search space should be explored, while at the same time the competence of the search engine is not compromised.  
	
	This study makes use of the $\epsilon$-greedy algorithms, which have been widely used in reinforcement learning situations, to guide the systematic exploration of the entire search space. By judiciously setting the value of $\epsilon$ in these algorithms, a good balance between exploitation and exploration may be achieved. We consider two variations of the $\epsilon$-greedy algorithm, representing different emphasis placed on fault-tolerance and efficiency, and study their performance both theoretically and experimentally. We have shown that, through such exploration, the problem of local optima can be overcome, and the algorithms are able to guarantee that the most relevant multimedia objects to given queries can always be found. Closed-form expressions of the performance of these algorithms have been derived, which exhibit good agreements with experimental results. These results show that such exploration paradigm may be usefully incorporated into multimedia search systems to enable them to enhance the performance of multimedia information search, so as to achieve the certain discovery of relevant objects that may be otherwise undiscoverable.


\begin{thebibliography}{10}
	\providecommand{\url}[1]{#1}
	\csname url@samestyle\endcsname
	\providecommand{\newblock}{\relax}
	\providecommand{\bibinfo}[2]{#2}
	\providecommand{\BIBentrySTDinterwordspacing}{\spaceskip=0pt\relax}
	\providecommand{\BIBentryALTinterwordstretchfactor}{4}
	\providecommand{\BIBentryALTinterwordspacing}{\spaceskip=\fontdimen2\font plus
		\BIBentryALTinterwordstretchfactor\fontdimen3\font minus
		\fontdimen4\font\relax}
	\providecommand{\BIBforeignlanguage}[2]{{%
			\expandafter\ifx\csname l@#1\endcsname\relax
			\typeout{** WARNING: IEEEtran.bst: No hyphenation pattern has been}%
			\typeout{** loaded for the language `#1'. Using the pattern for}%
			\typeout{** the default language instead.}%
			\else
			\language=\csname l@#1\endcsname
			\fi
			#2}}
	\providecommand{\BIBdecl}{\relax}
	\BIBdecl
	
	\bibitem{kofler2016user}
	C.~Kofler, M.~Larson, and A.~Hanjalic, ``User intent in multimedia search: a
	survey of the state of the art and future challenges,'' \emph{ACM Computing
		Surveys (CSUR)}, vol.~49, no.~2, p.~36, 2016.
	
	\bibitem{xie2019user}
	X.~Xie, ``User behavior modeling for web image search,'' in \emph{Proceedings
		of the Twelfth ACM International Conference on Web Search and Data
		Mining}.\hskip 1em plus 0.5em minus 0.4em\relax ACM, 2019, pp. 826--827.
	
	\bibitem{liem2017multimedia}
	C.~C. Liem, E.~G{\'o}mez, and G.~Tzanetakis, ``Multimedia technologies for
	enriched music performance, production, and consumption,'' \emph{IEEE
		MultiMedia}, no.~1, pp. 20--23, 2017.
	
	\bibitem{cao2017transitive}
	Z.~Cao, M.~Long, J.~Wang, and Q.~Yang, ``Transitive hashing network for
	heterogeneous multimedia retrieval,'' in \emph{Thirty-First AAAI Conference
		on Artificial Intelligence}, 2017.
	
	\bibitem{azzam2004implicit}
	I.~Azzam, C.~H. Leung, and J.~F. Horwood, ``Implicit concept-based image
	indexing and retrieval,'' in \emph{10th International Multimedia Modelling
		Conference, 2004. Proceedings.}\hskip 1em plus 0.5em minus 0.4em\relax IEEE,
	2004, pp. 354--359.
	
	\bibitem{stevenson2005comparative}
	K.~Stevenson and C.~Leung, ``Comparative evaluation of web image search engines
	for multimedia applications,'' in \emph{2005 IEEE International Conference on
		Multimedia and Expo}.\hskip 1em plus 0.5em minus 0.4em\relax IEEE, 2005,
	p.~4.
	
	\bibitem{datta2017multimodal}
	D.~Datta, S.~Varma, S.~K. Singh \emph{et~al.}, ``Multimodal retrieval using
	mutual information based textual query reformulation,'' \emph{Expert Systems
		with Applications}, vol.~68, pp. 81--92, 2017.
	
	\bibitem{leung2012intelligent}
	C.~H. Leung, A.~W. Chan, A.~Milani, J.~Liu, and Y.~Li, ``Intelligent social
	media indexing and sharing using an adaptive indexing search engine,''
	\emph{ACM Transactions on Intelligent Systems and Technology (TIST)}, vol.~3,
	no.~3, p.~47, 2012.
	
	\bibitem{fruit2018near}
	R.~Fruit, M.~Pirotta, and A.~Lazaric, ``Near optimal exploration-exploitation
	in non-communicating markov decision processes,'' in \emph{Advances in Neural
		Information Processing Systems}, 2018, pp. 2994--3004.
	
	\bibitem{narayanan2017event}
	V.~Narayanan and S.~Jagannathan, ``Event-triggered distributed control of
	nonlinear interconnected systems using online reinforcement learning with
	exploration,'' \emph{IEEE transactions on cybernetics}, vol.~48, no.~9, pp.
	2510--2519, 2017.
	
	\bibitem{vignesh2017online}
	N.~Vignesh and S.~Jagannathan, ``Online reinforcement with exploration for
	distributed control,'' in \emph{2017 International Joint Conference on Neural
		Networks (IJCNN)}.\hskip 1em plus 0.5em minus 0.4em\relax IEEE, 2017, pp.
	4022--4027.
	
	\bibitem{sutton2018reinforcement}
	R.~S. Sutton and A.~G. Barto, \emph{Reinforcement learning: An
		introduction}.\hskip 1em plus 0.5em minus 0.4em\relax MIT press, 2018.
	
	\bibitem{zamani2017relevance}
	H.~Zamani and W.~B. Croft, ``Relevance-based word embedding,'' in
	\emph{Proceedings of the 40th International ACM SIGIR Conference on Research
		and Development in Information Retrieval}.\hskip 1em plus 0.5em minus
	0.4em\relax ACM, 2017, pp. 505--514.
	
	\bibitem{peng2017overview}
	Y.~Peng, X.~Huang, and Y.~Zhao, ``An overview of cross-media retrieval:
	Concepts, methodologies, benchmarks, and challenges,'' \emph{IEEE
		Transactions on circuits and systems for video technology}, vol.~28, no.~9,
	pp. 2372--2385, 2017.
	
	\bibitem{Joachims2017}
	T.~Joachims, A.~Swaminathan, and T.~Schnabel, ``Unbiased learning-to-rank with
	biased feedback,'' in \emph{Proceedings of the Tenth ACM International
		Conference on Web Search and Data Mining}.\hskip 1em plus 0.5em minus
	0.4em\relax ACM, 2017, pp. 781--789.
	
	\bibitem{dang2017multimodal}
	D.-T. Dang-Nguyen, L.~Piras, G.~Giacinto, G.~Boato, and F.~G.~D. Natale,
	``Multimodal retrieval with diversification and relevance feedback for
	tourist attraction images,'' \emph{ACM Transactions on Multimedia Computing,
		Communications, and Applications (TOMM)}, vol.~13, no.~4, p.~49, 2017.
	
	\bibitem{over2004multimedia}
	P.~Over, C.~Leung, H.~Ip, and M.~Grubinger, ``Multimedia retrieval
	benchmarks,'' \emph{IEEE MultiMedia}, vol.~11, no.~2, pp. 80--84, 2004.
	
	\bibitem{sarwar2018term}
	S.~M. Sarwar, J.~Foley, and J.~Allan, ``Term relevance feedback for contextual
	named entity retrieval,'' in \emph{Proceedings of the 2018 Conference on
		Human Information Interaction \& Retrieval}.\hskip 1em plus 0.5em minus
	0.4em\relax ACM, 2018, p. 301.
	
	\bibitem{singhi2014cultural}
	A.~Singhi and D.~G. Brown, ``On cultural, textual and experiential aspects of
	music mood.'' in \emph{ISMIR}, 2014, pp. 3--8.
	
	\bibitem{kofler2015uploader}
	C.~Kofler, S.~Bhattacharya, M.~Larson, T.~Chen, A.~Hanjalic, and S.-F. Chang,
	``Uploader intent for online video: typology, inference, and applications,''
	\emph{IEEE transactions on multimedia}, vol.~17, no.~8, pp. 1200--1212, 2015.
	
	\bibitem{mei2014multimedia}
	T.~Mei, Y.~Rui, S.~Li, and Q.~Tian, ``Multimedia search reranking: A literature
	survey,'' \emph{ACM Computing Surveys (CSUR)}, vol.~46, no.~3, 2014.
	
	\bibitem{cui2014social}
	P.~Cui, S.-W. Liu, W.-W. Zhu, H.-B. Luan, T.-S. Chua, and S.-Q. Yang,
	``Social-sensed image search,'' \emph{ACM Transactions on Information Systems
		(TOIS)}, vol.~32, no.~2, p.~8, 2014.
	
	\bibitem{real2018regularized}
	E.~Real, A.~Aggarwal, Y.~Huang, and Q.~V. Le, ``Regularized evolution for image
	classifier architecture search,'' \emph{arXiv preprint arXiv:1802.01548},
	2018.
	
	\bibitem{liu2017hierarchical}
	H.~Liu, K.~Simonyan, O.~Vinyals, C.~Fernando, and K.~Kavukcuoglu,
	``Hierarchical representations for efficient architecture search,''
	\emph{arXiv preprint arXiv:1711.00436}, 2017.
	
	\bibitem{liu2018darts}
	H.~Liu, K.~Simonyan, and Y.~Yang, ``Darts: Differentiable architecture
	search,'' \emph{arXiv preprint arXiv:1806.09055}, 2018.
	
	\bibitem{rashedi2018comprehensive}
	E.~Rashedi, E.~Rashedi, and H.~Nezamabadi-pour, ``A comprehensive survey on
	gravitational search algorithm,'' \emph{Swarm and evolutionary computation},
	vol.~41, pp. 141--158, 2018.
	
	\bibitem{franzoni2017semantic}
	V.~Franzoni, Y.~Li, C.~H. Leung, and A.~Milani, ``Semantic evolutionary concept
	distances for effective information retrieval in query expansion,''
	\emph{arXiv preprint arXiv:1701.05311}, 2017.
	
	\bibitem{hirsch2017document}
	L.~Hirsch and A.~Di~Nuovo, ``Document clustering with evolved search queries,''
	in \emph{2017 IEEE Congress on Evolutionary Computation (CEC)}.\hskip 1em
	plus 0.5em minus 0.4em\relax IEEE, 2017, pp. 1239--1246.
	
	\bibitem{lu2014inferring}
	Z.~Lu, X.~Yang, W.~Lin, H.~Zha, and X.~Chen, ``Inferring user image-search
	goals under the implicit guidance of users,'' \emph{IEEE Transactions on
		Circuits and Systems For Video Technology}, vol.~24, no.~3, 2014.
	
	\bibitem{pan2014click}
	Y.~Pan, T.~Yao, T.~Mei, H.~Li, C.-W. Ngo, and Y.~Rui, ``Click-through-based
	cross-view learning for image search,'' in \emph{Proceedings of the 37th
		international ACM SIGIR conference on Research \& development in information
		retrieval}.\hskip 1em plus 0.5em minus 0.4em\relax ACM, 2014, pp. 717--726.
	
	\bibitem{gao2017learning}
	L.~Gao, J.~Song, X.~Liu, J.~Shao, J.~Liu, and J.~Shao, ``Learning in
	high-dimensional multimedia data: the state of the art,'' \emph{Multimedia
		Systems}, vol.~23, no.~3, pp. 303--313, 2017.
	
	\bibitem{zheng2018discriminatively}
	Z.~Zheng, L.~Zheng, and Y.~Yang, ``A discriminatively learned cnn embedding for
	person reidentification,'' \emph{ACM Transactions on Multimedia Computing,
		Communications, and Applications (TOMM)}, vol.~14, no.~1, p.~13, 2018.
	
	\bibitem{yang2015click}
	X.~Yang, Y.~Zhang, T.~Yao, C.-W. Ngo, and T.~Mei, ``Click-boosting
	multi-modality graph-based reranking for image search,'' \emph{Multimedia
		systems}, vol.~21, no.~2, pp. 217--227, 2015.
	
	\bibitem{li2018learning}
	K.~Li, G.-J. Qi, and K.~A. Hua, ``Learning label preserving binary codes for
	multimedia retrieval: A general approach,'' \emph{ACM Transactions on
		Multimedia Computing, Communications, and Applications (TOMM)}, vol.~14,
	no.~1, p.~2, 2018.
	
	\bibitem{zhang2008topological}
	H.~L. Zhang, C.~H. Leung, and G.~K. Raikundalia, ``Topological analysis of
	aocd-based agent networks and experimental results,'' \emph{Journal of
		Computer and System Sciences}, vol.~74, no.~2, pp. 255--278, 2008.
	
	\bibitem{ellis2014predicting}
	J.~G. Ellis, W.~S. Lin, C.-Y. Lin, and S.-F. Chang, ``Predicting evoked
	emotions in video,'' in \emph{2014 IEEE International Symposium on
		Multimedia}.\hskip 1em plus 0.5em minus 0.4em\relax IEEE, 2014, pp. 287--294.
	
	\bibitem{riegler2014reflects}
	M.~Riegler, M.~Larson, M.~Lux, and C.~Kofler, ``How'how'reflects what's what:
	content-based exploitation of how users frame social images,'' in
	\emph{Proceedings of the 22nd ACM international conference on
		Multimedia}.\hskip 1em plus 0.5em minus 0.4em\relax ACM, 2014, pp. 397--406.
	
	\bibitem{cobarzan2017interactive}
	C.~Cob{\^a}rzan, K.~Schoeffmann, W.~Bailer, W.~H{\"u}rst, A.~Bla{\v{z}}ek,
	J.~Loko{\v{c}}, S.~Vrochidis, K.~U. Barthel, and L.~Rossetto, ``Interactive
	video search tools: a detailed analysis of the video browser showdown 2015,''
	\emph{Multimedia Tools and Applications}, vol.~76, no.~4, pp. 5539--5571,
	2017.
	
	\bibitem{park2015large}
	J.~Y. Park, N.~O'Hare, R.~Schifanella, A.~Jaimes, and C.-W. Chung, ``A
	large-scale study of user image search behavior on the web,'' in
	\emph{Proceedings of the 33rd Annual ACM Conference on Human Factors in
		Computing Systems}.\hskip 1em plus 0.5em minus 0.4em\relax ACM, 2015, pp.
	985--994.
	
	\bibitem{feller2008introduction}
	W.~Feller, \emph{An introduction to probability theory and its
		applications}.\hskip 1em plus 0.5em minus 0.4em\relax John Wiley \& Sons,
	2008, vol.~1.
	
\end{thebibliography}


	\section*{Appendix A}
	From Section IV B, we have
	\begin{align*}
	f_{r,M,k+1} = f_{r,M,k} \nonumber \times \frac{{N-K-kr\choose r}}{{N-K-kr-1\choose r-1}} &\times \frac{{N-K-kr-1\choose r}}{{N-K-kr\choose r}}
	\\ &\times \frac{{N-K-(k+1)r-1\choose r-1}}{{N-K-(k+1)r\choose r}}, 
	\end{align*}
	
	This is expanded as
	\begin{align*}
	f_{r,M,k+1} = f_{r,M,k} \nonumber &\times \frac{[N-K-(k+1)r]!(r-1)!}{(N-K-kr-1)!} 
	\\ &\times \frac{(N-K-kr)!}{[N-K-(k+1)r]!r!}
	\\ &\times \frac{(N-K-kr-1)!}{[N-K-(k+1)r-1]!r!}
	\\ &\times \frac{[N-K-(k+1)r]!r!}{(N-K-kr)!}
	\\ &\times \frac{[N-K-(k+1)r-1]!}{[N-K-(k+2)r]!(r-1)!}
	\\ &\times \frac{[N-K-(k+2)r]!r!}{[N-K-(k+1)r]!}, 
	\end{align*}
	
	Considerable simplification of the above shows that $f_{r,M,k+1} = f_{r,M,k} = C$. Since after $(N-K)/r$ presentations, all random objects will have been exhausted, this implies that $C = r/(N-K)$. The mean is therefore 
	
	\begin{displaymath}
	C\left(1+2+ ...+\frac{(N-K)}{r}\right) = \frac{(N-M+2r)}{2r}.
	\end{displaymath}
	The corresponding second moment is 
	\begin{align}
	C\left(1^2+2^2+ ...+\big[\frac{(N-K)}{r}\big]^2\right) &= \bigg(1 + \frac{(N-K)}{r}\bigg) \nonumber
	\\ &\times \frac{\bigg(1 + \frac{2(N-K)}{r}\bigg)}{6}.
	\end{align}
	The variance is therefore
	\begin{displaymath}
	\bigg(1 + \frac{(N-K)}{r}\bigg) \times \frac{\bigg(1 + \frac{2(N-K)}{r}\bigg)}{6} - \bigg( \frac{(N-M+2r)}{2r}\bigg)^2
	\end{displaymath}
	which simplifies to 
	\begin{displaymath}
	\pushQED{\qed} 
	\frac{1}{12}\times\biggl\{\bigg[\frac{N-M+r}{r}\bigg]^2-1\biggr\}. \qedhere
	\popQED
	\end{displaymath}
\end{document}